# Non-gaussianity in axion N-flation models

# Quadratic and $\lambda\varphi^4$ plus axion potentials

Mehran Kamarpour

Email: M.kamarpour@sussex.ac.uk

Physics Department, College of Sciences, Shiraz University, Shiraz 71454, Iran

In this paper we investigate large non-gaussianity in axion N-flation models, taking account while dynamically a large number of axions begin away from the hilltop region(come down from the hill) and so serve only to be the source of the Hubble rate. Therefore the single field stays closest to the hilltop sources the non-Gaussianity. In this case most of axions can be replaced by a single effective field with a quadratic potential. So our potential will contain two fields. The full cosine is responsible for the axion closest to hilltop and quadratic term which is a source for Hubble rate [4]. We obtain power spectrum, spectral index and non-gaussianity parameter or $f_{nl}$ , then we impose conditions from WMAP for power spectrum and spectral index and see how large on $f_{nl}$ it is possible to achieve with such conditions. Finally we swap quadratic term to $\lambda\varphi^4$ and see whether this makes it harder or easier to achieve large $f_{nl}$.We find large non-gaussianity is achievable by imposing data from WMAP conditional on axion decay constant $f$ has reasonable value in connection with Planck mass and by requiring number of e-folds must be bounded between $40-60$.When we swap to $\lambda\varphi^4$ we find that it is harder to achieve non-Gaussianity , because we are imposed to investigate only $\lambda\varphi^4$ domination in consistency with WMAP data for spectral index. Although, in this case we find that large $f_{nl}$ is still achievable. Finally we verify imposing the condition $n-1\approx 0$ and find acceptable and detectable value for $f_{nl}$ typically of order of 10 and 100 depending on value of decay constant $f$.Swapping to $\lambda\varphi^4$ in this case does not give us any significant relation. In this paper we consider models which can generate large non-gausianity. We restrict ourselves to axion N-flation models.

**Contents**

## Introduction

It has been shown that inflation can solve some of the problems with pure hot big bang model. Inflation also generates perturbations by quantum fluctuations, which with slow-roll paradigm is expected to give close to scale –invariant Gaussian curvature perturbations. Also it has been shown that inflation can generate gravitational waves, with an amplitude which depends on the energy scale of inflation.[3].Over last ten years the observed power spectrum of primordial temperature fluctuations in the cosmic microwave background (WMAP)has provided evidence for inflation as the origin of cosmological fluctuations. Therefore we need to introduce and consider two-Field Models and assume slow-roll inflation. If we model inflation as being driven by canonically normalized slowly-rolling scalar fields, then we expect such fields generally produce Gaussian fluctuations at horizon exit. It is said that in single field inflation, the slow-roll conditions during which the seeds of the large scale are generated prevent the generation of observable non-gausianities. The reason is that the potential needs to be both flat enough for the fluctuations to develop and also steep enough for non-linearity to be significant[5].The increasingly accurate observations of the cosmic microwave background (WMAP and PLANCK) have great motivation of studying inflation beyond linear order in perturbation theory. Since there are many models which give similar predictions for spectral index and tensor to scalar ratio [6,7], it is important to consider models which give large non-Gausianities. In fact models with large non-Gausianities are powerful discriminant between different models of inflation.

It is mentioned above that single field inflation can't generate higher order perturbation during slow-roll conditions. The reason is that the form of potential which needs to be flat for the fluctuations to develop and steep enough for the non-linearity to be significant [5,9].So, large non-gaussianity is not achievable in this model. **In this paper** **we will see higher order perturbation is achievable in the case of single field inflation.**

## 1. Axion N-flation models

*The hypothesis of an invisible axion was made by Misha Shifman and others approximately thirty years ago. It has turned out to be an unusually fruitful idea crossing boundaries between particle physics ,astrophysics and cosmology*.[10]It was proposed to solve the so-called strong CP problem(existence of $\theta$ term in the action density of standard model of the elementary particles)[10].So, axion is the proposed particle which would set $\theta$ term to zero. This concept is now being used in many models of inflationary cosmology and dark matter. It is said, most of the matter density must be in the form of **nonbaryonic dark matter.** Essentially every known particle in the standard model has been ruled out to be a candidate for this dark matter. Fortunately, one of the plausible candidates beyond the standard model is axion [2].If we model inflation fields containing a large number of axions then we expect to generate large non-gaussianity of order 10 or 100 depending on what initial conditions are[11].The model is based on a set on $N_f$ uncoupled fields $\varphi_i$ each with potential

$$V_i = \Lambda_i^4 \left(1 - \cos\frac{2\pi\varphi_i}{f_i}\right) \qquad (1.1)$$

Where $f_i$ is the $i^{th}$ axion decay constant. The mass of each field in vacuum satisfies $m_i = \frac{2\pi\Lambda_i^2}{f_i}$ , and the angular field variables $\alpha_i = \frac{2\pi\varphi_i}{f_i}$ ,lie in the range $(-\pi, +\pi]$[11].Because the aim of introducing(invoking) this model is to find detectable non-gaussianity, we need to define non-gaussianity parameter..

### Non-gaussian perturbations

### Bispectrum

No additional correlation has been observed yet, for a cosmological perturbation described by linear perturbation theory [1].The problem is to define a three- point correlator which does not vanish like gaussian case. Its possible form in the non-gaussian case comes from reference [1] and is given by

$$\langle g_{k1} g_{k2} g_{k3} \rangle = (2\pi)^3 \delta^3_{k1+k2+k3} B_g(k_1, k_2, k_3) \qquad (1.2)$$

Where $B_g$ is called the **bispectrum.** *The delta function corresponds to invariance under translations, and the fact that the bispectrum depends only on the lengths of the three sides of the triangle formed by the momenta corresponds to invariance under rotations.*[1]

**Non-gaussianity**

From observations we know that curvature perturbation $\zeta$ is Gaussian to high accuracy,so the simplest possibility for the form of non-gaussianity is

$$\zeta(x) = \zeta_g + \zeta_{n.g.} \equiv g(x) + bg^2(x) \qquad (1.3)$$

Where g is a gaussian perturbation. This is known as local non-gaussianity[1].

**Prediction of the standard paradigm for non-gaussanity**

We need to go beyond the first order of curvature perturbation $\zeta$.We can write the comoving curvature perturbation during slow-roll inflation in the flat slicing (zero curvature gauges) as following relation.[1].Therefore we have

$$\zeta = -H\frac{\delta\varphi}{\dot{\varphi}} \qquad (1.4)$$

Using the slow-roll approximation $\dot{\varphi} = -\frac{V_{,\varphi}}{3H}$ by substituting into above equation we obtain

$$\zeta(x) = \frac{1}{M_{pl}^2}\frac{V}{V_{,\varphi}}\delta\varphi(x) \qquad (1.5)$$

For this purpose we use the $\delta N$ formula;

$$\zeta(x,t) = \delta N(x,t) \qquad (1.6)$$

$N$ at a given location depends only on $\varphi$.Remembering that we can expand $\delta N$ formula in terms of $\delta\varphi$.So we get

$$\zeta(x) = \delta N = N_{,\varphi}\delta\varphi(x) + \frac{1}{2}N_{,\varphi\varphi}(\delta\varphi(x))^2 + \cdots \qquad (1.7)$$

Now we return to our axion model.From equation (1.1) $V_i = \Lambda_i^4\left(1 - \cos\frac{2\pi\varphi_i}{f_i}\right)$ there are $N_f$ uncoupled inflation fields each with above potential. So, we need to calculate and derive required relations in order to obtain power spectrum, spectral index, tensor-scalar ratio and the non-gaussianity parameter $f_{nl}$ for this model. Generally $\delta N$ formula can be written [12]

$$\zeta = \delta N = \sum_i N_{,i}\delta\varphi_i + \frac{1}{2}\sum_{i,j} N_{,ij}\delta\varphi_i\delta\varphi_j + \frac{1}{6}\sum_{i,j,k} N_{i,j,k}\delta\varphi_i\delta\varphi_j\delta\varphi_k + \cdots \quad (1.8)$$

In general N-flation with $N_f$ uncoupled fields, the potential of inflation is given by

$$V = \sum_i^{N_f} V_i(\varphi_i) \quad (1.9)$$

Friedmann equation is given by

$$H^2 = \frac{1}{3M_{pl}^2}\sum_i^{N_f}\left(\frac{1}{2}\dot{\varphi}^2 + V_i(\varphi_i)\right) \quad (1.10)$$

And field equation will be

$$\ddot{\varphi}_i + 3H\dot{\varphi}_i = -V_{i,\varphi i} \quad (1.11)$$

Where $V_{i,\varphi i} = \partial V_i/\partial\varphi_i$

In slow-roll conditions we have

$$3M_{pl}^2H^2 \approx \sum_i^{N_f} V_i(\varphi_i) = V \quad (1.12)$$

And

$$3H\dot{\varphi}_i \approx -V_{i,\varphi i} \quad (1.13)$$

We define slow-roll parameters for each field like single field:

$$\epsilon_i = \frac{M_{pl}^2}{2}\left(\frac{V_{i,\varphi i}}{V_i}\right)^2 \quad (1.14)$$

From Eq.(1.12,13) by differentiating of both sides one can find

$$6M_{pl}^2H\dot{H} = \sum_i V_{i,\varphi i}\dot{\varphi}_i \rightarrow \dot{H} = \frac{1}{6HM_{pl}^2}\sum_i V_{i,\varphi i}\left(\frac{-V_{i,\varphi i}}{3H}\right) = -\frac{1}{6}\sum_i \frac{V_{i,\varphi i}^2}{3M_{pl}^2H^2}$$

But $3M_{pl}^2H^2 \approx V$ so we will get

$$\dot{H} = -\frac{1}{6}\frac{\sum_i^{N_f} V_{i,\varphi i}^2}{V} \qquad (1.15)$$

The global slow-roll parameter $\epsilon_H$is related to $\epsilon_i$ by following relation:

$$\epsilon_H = -\frac{\dot{H}}{H^2} = -\frac{-\frac{1}{6}\frac{\sum_i V_{i,\varphi i}^2}{V}}{\frac{V}{3M_{pl}^2}}$$

After simplification what we obtain is

$$\epsilon_H = \frac{1}{2}M_{pl}^2\frac{\sum_i V_{i,\varphi i}^2}{V^2} \qquad (1.16)$$

If we substitute $\epsilon_i = \frac{M_{pl}^2}{2}\left(\frac{V_{i,\varphi i}}{V_i}\right)^2$into above equation we find

$$\epsilon_H = \sum_i \left(\frac{V_i}{V}\right)^2 \epsilon_i \qquad (1.17)$$

The above relation says each field contributes according to its share of the total energy density. During inflation we must have $\epsilon_H < 1$.

**Number of e-folds**

Number of e-folds in this case is given by [12]

$$N = \frac{1}{M_{pl}^2}\sum_i^{N_f}\int_{\varphi_i^{end}}^{\varphi_i}\frac{V_i}{V_{i,\varphi i}}d\varphi_i \qquad (1.18)$$

Therefore we obtain

$$\partial N/\partial\varphi_i = N_{,i} = \frac{1}{M_{pl}^2}\frac{V_i}{V_{i,\varphi i}} \qquad (1.19)$$

$$N_{,ij} = \frac{\partial}{\partial\varphi_j}\left(\frac{\partial N}{\partial\varphi_i}\right) = \frac{1}{M_{pl}^2}\frac{\partial}{\partial\varphi_j}\left(\frac{V_i}{V_{i,\varphi i}}\right) = \frac{1}{M_{pl}^2}\frac{\frac{\partial V_i}{\partial\varphi_j}V_{i,\varphi i} - \frac{\partial V_{i,\varphi i}}{\partial\varphi_j}V_i}{\left(V_{i,\varphi i}\right)^2} \to$$

$$= \frac{1}{M_{pl}^2}\frac{\partial}{\partial\varphi_j}\left(\frac{V_i}{V_{i,\varphi i}}\right) = \frac{1}{M_{pl}^2}\frac{\frac{\partial V_i}{\partial\varphi_i}\frac{\partial\varphi_i}{\partial\varphi_j}V_{i,\varphi i} - \frac{\partial V_{i,\varphi i}}{\partial\varphi_i}\frac{\partial\varphi_i}{\partial\varphi_j}V_i}{\left(V_{i,\varphi i}\right)^2} \to$$

$$= \frac{1}{M_{pl}^2} \frac{\partial}{\partial \varphi_j} \left( \frac{V_i}{V_{i,\varphi i}} \right) = \frac{1}{M_{pl}^2} \frac{\frac{\partial V_i}{\partial \varphi_i} \delta_{ij} V_{i,\varphi i} - \frac{\partial V_{i,\varphi i}}{\partial \varphi_i} \delta_{ij} V_i}{\left(V_{i,\varphi i}\right)^2} = \frac{1}{M_{pl}^2} \frac{\left({V_{i,\varphi i}}^2 - V_{i,\varphi i \varphi i} V_i\right) \delta_{ij}}{\left(V_{i,\varphi i}\right)^2} \rightarrow$$

$$N_{,ij} = \frac{1}{M_{pl}^2} \left( 1 - \frac{V_{i,\varphi i \varphi i} V_i}{{V_{i,\varphi i}}^2} \right) \delta_{ij} \qquad (1.20)$$

Note that $V_{i,\varphi i \varphi i} = \frac{\partial}{\partial \varphi_i} \left( \frac{\partial V_i}{\partial \varphi_i} \right)$.

**Power spectrum**

Using $\delta N$ formalism, the power spectrum of scalar perturbation is given by [1,11]

$$p_\zeta = \frac{{H_*}^2}{4\pi^2} \sum_i N_{,i} N_{,i} \qquad (1.21)$$

Note that $*$ indicates evaluation at horizon crossing. From equation (1.19) $\partial N / \partial \varphi_i = N_{,i} = \frac{1}{M_{pl}^2} \frac{V_i}{V_{i,\varphi i}}$ and equation (1.1) $V_i = \Lambda_i^4 \left( 1 - \cos \frac{2\pi \varphi_i}{f_i} \right)$ one can find

$$N_{,i} = \frac{1}{M_{pl}^2} \frac{\Lambda_i^4 \left( 1 - \cos \frac{2\pi \varphi_i}{f_i} \right)}{\frac{2\pi}{f_i} \Lambda_i^4 \sin \frac{2\pi \varphi_i}{f_i}} = \frac{f_i}{2\pi M_{pl}^2} \frac{1 - \cos \frac{2\pi \varphi_i}{f_i}}{\sin \frac{2\pi \varphi_i}{f_i}} \qquad (1.22)$$

From equation (1.14) $\epsilon_i = \frac{M_{pl}^2}{2} \left( \frac{V_{i,\varphi i}}{V_i} \right)^2$ we have

$$\epsilon_i = \frac{2 M_{pl}^2 \pi^2}{{f_i}^2} \left( \frac{1 + \cos \frac{2\pi \varphi_i}{f_i}}{1 - \cos \frac{2\pi \varphi_i}{f_i}} \right) \qquad (1.23)$$

Combining $\epsilon_i$ $and$ $N_{,i}$ and plugging into $p_\zeta$ we finally find following relation.

$$p_\zeta = \frac{{H_*}^2}{8\pi^2 M_{pl}^2} \sum_i \frac{1}{\epsilon_i^*} \qquad (1.24)$$

**Spectral index and tensor-scalar ratio**

Spectral index is given by [1,3,13]

$$n-1=\frac{d\ln p_\zeta}{d\ln k}=\frac{1}{H_*}\frac{d\ln p_\zeta}{dt}$$

$$=-2\epsilon_H^*-\frac{8\pi^2}{3{H_*}^2}\sum_i\frac{{\Lambda_j}^4}{{f_j}^2}\frac{1}{\epsilon_j^*}\Bigg/\sum_i\frac{1}{\epsilon_i^*} \tag{1.25}$$

From references [3,11,13] tensor to scalar ratio is given by

$$r=\frac{p_T}{p_\zeta}=16\Bigg/\sum_i\frac{1}{\epsilon_i^*} \tag{1.26}$$

**Non-gaussianity parameter $f_{nl}$**

$f_{nl}$ Can be written by following relation [11].We have

$$f_{nl}=\frac{5}{6}\frac{\sum_{ij}N_{,i}N_{,j}N_{,ij}}{\left(\sum_k N_{,k}N_{,k}\right)^2} \tag{1.27}$$

From equation (1.20) we have $N_{,ij}=\frac{1}{M_{pl}^2}\left(1-\frac{V_{i,\varphi i\varphi i}V_i}{{V_{i,\varphi i}}^2}\right)\delta_{ij}$ so, $N_{,ij}\propto\delta_{ij}$

Just we need to find $N_{,ii}=\frac{\partial}{\partial\varphi_i}\left(\frac{\partial N}{\partial\varphi_i}\right)$.After a little algebra, we find

$$N_{,ii}=\left(1-\frac{V_{i,\varphi i\varphi i}V_i}{{V_{i,\varphi i}}^2}\right)=\frac{1}{M_{pl}^2}\frac{1}{1+\cos\frac{2\pi\varphi_i}{f_i}} \tag{1.28}$$

If we compare equations (1.22) and (1.23) we can find following relation between two relations:

$$\left(N_{,i}\right)^2=\frac{f_i^2}{(2\pi)^2M_{pl}^4}\frac{1-\cos\frac{2\pi\varphi_i}{f_i}}{1+\cos\frac{2\pi\varphi_i}{f_i}}=\frac{1}{2M_{pl}^2}\frac{1}{\epsilon_i} \tag{1.29}$$

So what we obtain for $f_{nl}$is,

$$\frac{6}{5}f_{nl}=\frac{2\sum_i\frac{1}{\epsilon_i^*}\frac{1}{1+\cos\alpha_i^*}}{\left(\sum_i\frac{1}{\epsilon_i^*}\right)^2}\rightarrow, r=16\Bigg/\sum_i\frac{1}{\epsilon_i^*}\rightarrow$$

$$\frac{6}{5}f_{nl}=\frac{r^2}{128}\sum_i\frac{1}{\epsilon_i^*}\frac{1}{1+\cos\alpha_i^*} \tag{1.30}$$

$\alpha_i^* = \frac{2\pi\varphi_i^*}{f_i}$. $*$Denotes evaluation at horizon crossing.

**How many inflations stay around the hilltop?**

We are now in a position to obtain a relation for number of inflations which stay around the hilltop. It can be shown that a few inflations stays around the hilltop.

The $\epsilon_i$ approach zero for fields close to the hilltop, so each summation in Eqs.(1.24)-(1.26) and (1.30)is dominated by those fields by smallest $\epsilon_i$ .Suposse some number $\bar{N}$ of fields have roughly comparable $\epsilon_i$ ,of order $\bar{\epsilon}$ [11],which is related to angle $\bar{\alpha}$ by following relations;

$$\epsilon_i = \frac{2M_{pl}^2\pi^2}{{f_i}^2}\left(\frac{1+\cos\alpha_i}{1-\cos\alpha_i}\right) \tag{1.23}$$

So we can rewrite this equation in terms of $\bar{\alpha}$ we obtain

$$\bar{\epsilon} = \frac{2\pi^2M_{pl}^2}{f^2}\frac{1+\cos\bar{\alpha}}{1-\cos\bar{\alpha}} = \frac{2\pi^2M_{pl}^2}{f^2}\frac{1+\cos\bar{\alpha}}{1+(\cos(\pi-\bar{\alpha}))} \approx \frac{2\pi^2M_{pl}^2}{f^2}\frac{1+\cos\bar{\alpha}}{1+\left[1-\left(\frac{\pi-\bar{\alpha}}{2}\right)^2\right]} \rightarrow$$

Note that around the hilltop $(\pi-\bar{\alpha}) \ll 1$ and so we can find other relations:

$$\bar{\epsilon} \approx \frac{\pi^2M_{pl}^2}{f^2}(1-\cos(\pi-\bar{\alpha})) \approx \frac{\pi^2M_{pl}^2}{2f^2}(\pi-\bar{\alpha})^2 \tag{1.31}$$

.

$$\sum\nolimits_i \frac{1}{\epsilon_i^*} \approx \frac{\bar{N}}{\bar{\epsilon}} \tag{1.32}$$

Plugging into equation (1.27) we find

$$p_\zeta = \frac{{H_*}^2}{8\pi^2M_{pl}^2}\sum_i \frac{1}{\epsilon_i} \approx \frac{H_*^2}{8\pi^2M_{pl}^2}\frac{\bar{N}}{\bar{\epsilon}} \tag{1.33}$$

Substituting $\bar{\epsilon}$ from equation (1.31) we obtain

$$p_\zeta \approx \frac{H^2\bar{N}f^2}{4\pi^4M_{pl}^4(\pi-\bar{\alpha})^2} \tag{1.34}$$

Tensor-scalar ratio becomes

$$r = \frac{p_T}{p_\zeta} = 16\Big/\sum_i \frac{1}{\epsilon_i^*} \approx 16\frac{\bar{\epsilon}}{\bar{N}} \approx \frac{8\pi^2M_{pl}^2(\pi-\bar{\alpha})^2}{\bar{N}f^2} \tag{1.35}$$

**Behaviour of $f_{nl}$ around the hilltop**

If we consider equation (1.30) $\frac{6}{5}f_{nl}=\frac{r^2}{128}\sum_i\frac{1}{\epsilon_i^*}\frac{1}{1+\cos\alpha_i^*}$ and substitute for $\alpha_i^*$ , $\epsilon_i^*$ and $r^2$ from above relations then we can find

$$\frac{6}{5}f_{nl}\approx\frac{r^2}{128}\frac{\bar{N}}{\bar{\epsilon}}\frac{1}{1-\cos(\pi-\bar{\alpha})}\rightarrow$$

Expanding cosine function and remembering that $(\pi-\bar{\alpha})\ll 1$ we will find

$$\frac{6}{5}f_{nl}\approx\frac{2\pi^2}{\bar{N}}\left(\frac{M_p}{f}\right)^2 \qquad (1.36)$$

**This relation , exactly is the relation (9) of authors in reference [11] and shows the approximate behaviour of $f_{nl}$ which is independent of $\bar{\epsilon}$ if the dominant fields are sufficiently close to hilltop**. If we consider $\bar{N}\geq 1$ then

$$f_{nl}\lesssim\frac{5\pi^2}{3}\left(\frac{M_p}{f}\right)^2 \qquad (1.37)$$

If we require the classical motion of inflation $\bar{\varphi}=f\bar{\alpha}/2\pi$ per Hubble time can't be less than its quantum fluctuations $H/2\pi$, then we need

$$\varphi_{hiltop}-\bar{\varphi}\approx H/2\pi\,,\varphi_{hiltop}-\bar{\varphi}\gtrsim H/2\pi \qquad (1.38)$$

Then we find

$$\frac{f}{2}-\frac{f\bar{\alpha}}{2\pi}\gtrsim\frac{H}{2\pi}\rightarrow\frac{f(\pi-\bar{\alpha})}{2\pi}\gtrsim\frac{H}{2\pi}\rightarrow$$

$$\boxed{(\pi-\bar{\alpha})\gtrsim\frac{H}{f}} \qquad (1.39)$$

Note that $\varphi_{Hilltop}=\frac{f\alpha_{Hilltop}}{2\pi}=\frac{f\pi}{2\pi}=\frac{f}{2}$.This relation will be used soon when we obtain number of inflations which stay around the hilltop. From Eq.(1.34) $p_\zeta\approx\frac{H^2\bar{N}f^2}{4\pi^4M_{pl}^4(\pi-\bar{\alpha})^2}$ we can get

$$\boxed{(\pi-\bar{\alpha})\approx\frac{H\sqrt{\bar{N}}f}{2\pi^2M_{pl}^2\sqrt{p_\zeta}}} \qquad (1.40)$$

Comparing (1.39) and (1.40) we have

$$\boxed{\sqrt{\bar{N}}f^2 \gtrsim 2\pi^2 M_{pl}^2\sqrt{p_\zeta}} \qquad (1.41)$$

From (1.36) we have $\frac{6}{5}f_{nl} \approx \frac{2\pi^2}{\bar{N}}\left(\frac{M_p}{f}\right)^2$.This relation can be used for $\bar{N}$.It follows

$$\boxed{\sqrt{\bar{N}} \lesssim \frac{5}{6}\frac{1}{f_{nl}\sqrt{p_\zeta}}} \qquad (1.42)$$

WMAP observations requires $p_\zeta \approx 2 \times 10^{-9}$ [8], and then

$$\sqrt{\bar{N}} \lesssim \frac{1.86\ \times 10^4}{f_{nl}} \qquad (1.43)$$

Finally, there is a relation between $\bar{N}$ and $f_{nl}$ as follows.

$$\bar{N} \lesssim \frac{25}{36}\frac{1}{p_\zeta}\frac{1}{(f_{nl})^2} \qquad (1.44)$$

What we consider above, about the condition in relation (1.39) is different from the author of reference [13] who considers another condition. The condition is $(\pi - \bar{\alpha}) \gtrsim \frac{3fH^3}{4\pi^2\Lambda^4} = 2\xi^{-1}\frac{H}{f}$ where $\xi \ll 1$.

**Non-gaussianities in two-field inflation**

Now we assume most of axions can be replaced by a single effective field with a quadratic model. The result is ,two separable potentials. In this case we compute power spectrum, spectral index and non-gaussianity parameter $f_{nl}$.We investigate the model by data from WMAP and see generally and numerically how large on $f_{nl}$ is it possible to achieve?

## 2. The Model

The model is based on relation (5.2) page 19 of reference[4] and is given by following relation

$$W = \frac{1}{2}m^2\varphi^2 + \Lambda^4\left(1 - \cos\frac{2\pi x}{f}\right) \qquad (2.1)$$

In fact the model contains two potentials, including one axion on(closest) hilltop and a quadratic term. So the model obeys general form of separable into the sum of two potentials each of which is dependent on only one of the two fields.

$$W(\varphi, x) = U(\varphi) + V(x) \qquad (2.2)$$

**Slow-roll dynamics**

The equations for the background fields read [14]

$$\ddot{\varphi} + 3H\dot{\varphi} + U_{,\varphi} = 0 \qquad (2.3)$$

$$\ddot{x} + 3H\dot{x} + V_{,x} = 0 \qquad (2.4)$$

Where

$$U_{,\varphi} \equiv \frac{dU}{d\varphi}\,, V_{,x} = \frac{dV}{dx} \qquad (2.5)$$

The unperturbed Friedmann equations read

$$H^2 = \frac{1}{3M_{pl}^2}\left(\frac{1}{2}\dot{\varphi}^2 + \frac{1}{2}\dot{x}^2 + W\right) \qquad (2.6)$$

$$\dot{H} = -\frac{1}{2M_{pl}^2}(\dot{\varphi}^2 + \dot{x}^2) \qquad (2.7)$$

Inflation takes place when $\epsilon_H = -H/\dot{H}^2 < 1$.Under the slow-roll conditions field equations reduce to

$$3H\dot{\varphi} \approx -U_{,\varphi}\,, \quad 3H\dot{x} \approx -V_{,x}\,, \quad 3M_{pl}^2 H^2 \approx W \qquad (2.8)$$

Slow-roll parameters are

$$\epsilon_\varphi = \frac{M_{pl}^2}{2}\left(\frac{U_{,\varphi}}{U}\right)^2, \epsilon_x = \frac{M_{pl}^2}{2}\left(\frac{V_{,x}}{V}\right)^2 \qquad (2.9)$$

**Number of e-folds**

From reference[12] Number of e-folds is defined to be

$$N = \frac{1}{M_{pl}^2}\int_{\varphi_{end}}^{\varphi_*}\frac{U}{U_{,\varphi}}d\varphi + \frac{1}{M_{pl}^2}\int_{x_{end}}^{x_*}\frac{V}{V_{,x}}dx \qquad (2.10)$$

All we need to do is to compute required relations from our model (2.1).Therefore,

$$U(\varphi) = \frac{1}{2}m^2\varphi^2\,, V(x) = \Lambda^4\left(1 - \cos\frac{2\pi x}{f}\right) \qquad (2.11)$$

$$\epsilon_\varphi = \frac{{M_{pl}}^2}{2}\left(\frac{U_{,\varphi}}{U}\right)^2 = \frac{2M_{pl}^2}{\varphi^2} \tag{2.12}$$

$$\epsilon_x = \frac{{M_{pl}}^2}{2}\left(\frac{V_{,x}}{V}\right)^2 = \frac{2M_{pl}^2\pi^2}{f^2}\left[\frac{1+\cos\frac{2\pi x}{f}}{1-\cos\frac{2\pi x}{f}}\right] \tag{2.13}$$

From equation (2.10) one can find

$$N_{,\varphi} = \frac{\partial N}{\partial \varphi} = \frac{1}{M_{pl}^2}\frac{U}{U_{,\varphi}} = \frac{\varphi}{2M_{pl}^2} = \frac{1}{M_{pl}}\frac{1}{\sqrt{2\epsilon_\varphi}} \tag{2.14}$$

$$N_{,\varphi\varphi} = \frac{\partial}{\partial\varphi}\left(\frac{\partial N}{\partial\varphi}\right) = \frac{1}{2M_{pl}^2} \tag{2.15}$$

$$N_{,x} = \frac{\partial N}{\partial x} = \frac{1}{M_{pl}^2}\frac{V}{V_{,x}} = \frac{f}{2\pi M_{pl}^2}\frac{1-\cos\frac{2\pi x}{f}}{\sin\frac{2\pi x}{f}}$$

$$= \frac{1}{M_{pl}}\frac{1}{\sqrt{2\epsilon_x}} \tag{2.16}$$

$$N_{,xx} = \frac{\partial}{\partial x}\left(\frac{\partial N}{\partial x}\right) = \frac{1}{M_{pl}^2}\frac{1}{1+\cos\frac{2\pi x}{f}} \tag{2.17}$$

$${N_{,x}}^2 = \frac{f^2}{4\pi^2 M_{pl}^4}\frac{1-\cos\frac{2\pi x}{f}}{1+\cos\frac{2\pi x}{f}} = \frac{1}{2M_{pl}^2}\frac{1}{\epsilon_x} \tag{2.18}$$

$${N_{,\varphi}}^2 = \frac{\varphi^2}{4{M_{pl}}^4} = \frac{1}{2M_{pl}^2}\frac{1}{\epsilon_\varphi} \tag{2.19}$$

We know that $N_{,x\varphi} = N_{,\varphi x} = 0$ so, we are ready to calculate power spectrum and other relevant observables.

**Power spectrum for sum of two separable potentials**

From equation (1.21) we have

$$p_\zeta = \frac{{H_*}^2}{4\pi^2}\sum_i N_{,i}N_{,i} = \frac{{H_*}^2}{4\pi^2}\left[{N_{,x}}^2 + {N_{,\varphi}}^2\right] \tag{2.20}$$

Substituting ${N_{,x}}^2, {N_{,\varphi}}^2$ from Eqs.(2.18) and (2.19) power spectrum is found to be

$$p_\zeta = \frac{{H_*}^2}{4\pi^2}\left[\frac{f^2}{4\pi^2{M_{pl}}^4}\frac{1-\cos\left(\frac{2\pi x}{f}\right)}{1+\cos\left(\frac{2\pi x}{f}\right)} + \frac{\varphi^2}{4{M_{pl}}^4}\right] \tag{2.21}$$

Using $\epsilon_x, \epsilon_\varphi$ from equations (2.12) and (2.13) we can write power spectrum in terms of slow-roll parameters. It follows

$$p_\zeta = \frac{{H_*}^2}{8\pi^2 {M_{pl}}^2}\left[\frac{1}{\epsilon_x} + \frac{1}{\epsilon_\varphi}\right] \tag{2.22}$$

Where $*$ denotes evaluation at horizon crossing. From now on $*$ denotes horizon crossing otherwise specified. Slow-roll conditions (2.8) implies that $p_\zeta$ can be written as

$$p_\zeta = \frac{W_*}{24\pi^2 {M_{pl}}^4}\left[\frac{1}{\epsilon_x} + \frac{1}{\epsilon_\varphi}\right] \tag{2.23}$$

**Spectral index in two separable inflation fields**

In order to compute spectral index one needs to differentiate $\ln p_\zeta$ with respect to time, but one just needs to be careful about evaluation at horizon crossing. Therefore, one finds

$$n - 1 = \frac{d \ln p_\zeta}{d \ln k} = \frac{1}{H_*}\frac{d \ln p_\zeta}{dt} = \frac{1}{H_*}\frac{d}{dt}\left\{\ln\left[\frac{{H_*}^2}{8\pi^2 {M_{pl}}^2}\left(\frac{1}{\epsilon_x} + \frac{1}{\epsilon_\varphi}\right)\right]\right\}$$

$$= \frac{1}{H_*}\frac{d}{dt}\left\{\ln\frac{{H_*}^2}{8\pi^2 {M_{pl}}^2} + \ln\left(\frac{1}{\epsilon_x} + \frac{1}{\epsilon_\varphi}\right)\right\} \rightarrow$$

$$= \frac{2\dot{H}}{{H_*}^2} - \frac{1}{3{H_*}^2}\left[\frac{\frac{2m^2}{\epsilon_\varphi} + \frac{8\pi^2 \Lambda^4}{f^2}\frac{1}{\epsilon_x}}{\frac{1}{\epsilon_\varphi} + \frac{1}{\epsilon_x}}\right] \rightarrow$$

$$n - 1 = -2\epsilon_H - \frac{1}{3{H_*}^2}\left[\frac{\frac{2m^2}{\epsilon_\varphi} + \frac{8\pi^2 \Lambda^4}{f^2}\frac{1}{\epsilon_x}}{\frac{1}{\epsilon_\varphi} + \frac{1}{\epsilon_x}}\right] \tag{2.24}$$

If we substitute $\epsilon_x, \epsilon_\varphi$ then we can obtain a relation in terms of fields $x, \varphi$ which is useful for numerical calculations. We find

$$n-1=-2\epsilon_H-\frac{1}{3{H_*}^2}\left\{\left[\left(\frac{\frac{m^2\varphi^2}{{M_{pl}}^2}}{\frac{\varphi^2}{2{M_{pl}}^2}+\frac{f^2}{2\pi^2{M_{pl}}^2}\left(\frac{1-\cos\frac{2\pi x}{f}}{1+\cos\frac{2\pi x}{f}}\right)}\right)\right]\right.$$

$$\left.+\left[\frac{\frac{4\Lambda^4}{{M_{pl}}^2}\left(\frac{1-\cos\frac{2\pi x}{f}}{1+\cos\frac{2\pi x}{f}}\right)}{\frac{\varphi^2}{2{M_{pl}}^2}+\frac{f^2}{2\pi^2{M_{pl}}^2}\left(\frac{1-\cos\frac{2\pi x}{f}}{1+\cos\frac{2\pi x}{f}}\right)}\right]\right\}$$

(2.25)

Note that $\epsilon_H=-\frac{\dot{H}}{H^2}=\frac{U^2\epsilon_\varphi+V^2\epsilon_X}{W^2}$ (See equation (1.17).So, $\epsilon_H$ is found by following relation

$$\epsilon_H=\frac{\frac{m^4\varphi^2{M_{pl}}^2+}{2}+\frac{f^2m^4{M_{pl}}^2\left(\sin\frac{2\pi x}{f}\right)^2}{8\pi^2}}{\left(\frac{1}{2}m^2\varphi^2+\Lambda^4\left(1-\cos\frac{2\pi x}{f}\right)\right)^2} \qquad (2.26)$$

Note that $\Lambda^4=f^2m^2/4\pi^2$. We will return to this formula when we investigate numerical data from WMAP.

It is useful to rewrite equation (2.25) then one can estimate numerical calculations for spectral index. Using slow-roll equation for ${H_*}^2$and substituting $\Lambda^4=f^2m^2/4\pi^2$ into equation (2.25) then one obtains

$$n-1$$

$$=-\frac{\left[m^4\varphi^2{M_{pl}}^2+\frac{f^2m^4{M_{pl}}^2}{4\pi^2}\left(\sin\frac{2\pi x}{f}\right)^2\right]}{\left[\frac{1}{2}m^2\varphi^2+\frac{f^2m^2}{4\pi^2}\left(1-\cos\frac{2\pi x}{f}\right)\right]^2}$$

$$-\frac{\left[\frac{f^2m^2}{\pi^2}\left(\frac{\left(\sin\frac{2\pi x}{f}\right)^2}{\left(1+\cos\frac{2\pi x}{f}\right)^2}\right)+m^2\varphi^2\right]}{\left[\frac{1}{2}m^2\varphi^2+\frac{f^2m^2}{4\pi^2}\left(1-\cos\frac{2\pi x}{f}\right)\right]}\left(\frac{1}{\frac{f^2}{2\pi^2{M_{pl}}^2}\left(\frac{1-\cos\frac{2\pi x}{f}}{1+\cos\frac{2\pi x}{f}}\right)+\frac{\varphi^2}{2{M_{pl}}^2}}\right)$$

(2.25)´

**Tensor-scalar ratio**

From equation (1.26) one can obtain

$$r = \frac{p_T}{p_\zeta} = 16 \Big/ \sum_i \frac{1}{\epsilon_i^*} = \frac{16}{\frac{1}{\epsilon_\varphi} + \frac{1}{\epsilon_x}} \tag{2.27}$$

**Calculation of $f_{nl}$.**

Equation (1.27) gives us the amount of non-linearity. If we compute $f_{nl}$from that relation we get

$$f_{nl} \approx \frac{5}{6} \frac{\sum_{ij} N_{,i} N_{,j} N_{,ij}}{\left(\sum_k N_{,k} N_{,k}\right)^2} = \frac{5}{6} \frac{{N_{,\varphi}}^2 N_{,\varphi\varphi} + {N_{,x}}^2 N_{,xx}}{\left({N_{,\varphi}}^2 + {N_{,x}}^2\right)^2} \tag{2.28}$$

Substituting $N_{,\varphi}, N_{,x}, N_{,\varphi\varphi}, N_{,xx}$ from equations (2.14-2.19) $f_{nl}$is given by

$$\frac{6}{5} f_{nl} \approx \frac{\frac{1}{\epsilon_\varphi} + \frac{2}{\epsilon_x}\left(\frac{1}{1 + \cos\frac{2\pi x}{f}}\right)}{\left(\frac{1}{\epsilon_\varphi} + \frac{1}{\epsilon_x}\right)^2} \tag{2.29}$$

This relation can be written in terms of inflation fields $x, \varphi$ which is very useful for numerical calculations. In such case we have (see eqs.2.12 and 2.13).

$$\epsilon_\varphi = \frac{2M_{pl}^2}{\varphi^2}, \epsilon_x = \frac{1}{M_{pl}^2}\left(\frac{V_{,x}}{V}\right)^2 = \frac{2M_{pl}^2\pi^2}{f^2}\left[\frac{1 + \cos\frac{2\pi x}{f}}{1 - \cos\frac{2\pi x}{f}}\right] \rightarrow$$

$$\frac{6}{5} f_{nl} \approx \frac{\frac{\varphi^2}{2{M_{pl}}^2} + \frac{f^2}{{M_{pl}}^2\pi^2}\left[\frac{1 - \cos\frac{2\pi x}{f}}{1 + \cos\frac{2\pi x}{f}}\right]\left[\frac{1}{1 + \cos\frac{2\pi x}{f}}\right]}{\left[\frac{\varphi^2}{2{M_{pl}}^2} + \frac{f^2}{2{M_{pl}}^2\pi^2}\left(\frac{1 - \cos\frac{2\pi x}{f}}{1 + \cos\frac{2\pi x}{f}}\right)\right]^2} \tag{2.30}$$

Note that $\Lambda^4 = f^2 m^2/4\pi^2$.Also there is another way of writing $f_{nl}$as follows.

Remembering equation (2.27) $r = \frac{p_T}{p_\zeta} = 16/\sum_i \frac{1}{\epsilon_i^*} = \frac{16}{\frac{1}{\epsilon_\varphi} + \frac{1}{\epsilon_x}}$

Substituting $r^2$ into (2.29) So , $f_{nl}$ can be written as

$$\frac{6}{5}f_{nl} \approx \frac{r^2}{128}\left[\frac{1}{2\epsilon_\varphi}+\frac{1}{\epsilon_x}\frac{1}{1+\cos\frac{2\pi x}{f}}\right] \qquad (2.31)$$

**NUMERICAL CALCULATIONS**

Now by imposing data we investigate how large $f_{nl}$ is it possible? As a result how large $f_{nl}$ can be found in various cases considering constrains on spectral index and power spectrum by WMAP. However, one can assume that spectral index is nearly zero and try to obtain a relation for $f_{nl}$.Here for two choices of initial condition we investigate values of $f_{nl}$.At first one can invoke equation (2.10) for $N$ in order to obtain a relation for number of e-folds and then insert initial conditions for $\varphi_*, x_*$ $and$ $N$.For instance for $\varphi_* = 14M_{pl}, x_* = (1/2 - 0.001/2\pi)f$ , $N = 50$ we find $f^2 = 5.194736842{M_{pl}}^2$ and $n = 0.959322092$ and $f_{nl} \approx 3.166077895$.Note that $\alpha = \frac{2\pi x_*}{f} = \pi - 0.001$ for axion field. The results are outlined in table below. **Note that we ignore very small correction from the location of end of inflation $\boldsymbol{x_{end}}$**.Just in one case we take into account contribution of this location for axion filed in tables (1) and (2).There is no difference in value of spectral index. But there is a small correction on $f_{nl}$.However ignoring $x_{end}$ in numerical calculations is not very important because we are not using exact relation for $f_{nl}$.

Table (1)

| $N$ | $\varphi_*$ | $x_*$ | $\alpha_*$ | $f^2$ | $n$ | $f_{nl}$ |
|---|---|---|---|---|---|---|
| 50 | $14M_{pl}$ | $(1/2 - 0.001/2\pi)f$ | $\pi - 0.001$ | $5.20{M_{pl}}^2$ | 0.959 | 3.17 |
| 60 | $14M_{pl}$ | $(1/2 - 0.001/2\pi)f$ | $\pi - 0.001$ | $29.87{M_{pl}}^2$ | 0.959 | 0.661 |
| 48.51 | $14M_{pl}$ | $(1/2 - 0.001/2\pi)f$ | $\pi - 0.001$ | $0.026{M_{pl}}^2$ | 0.959 | 607 |
| 48.51 | $14M_{pl}$ | $(1/2 - 0.001/2\pi)f$ | $\pi - 0.001$ | 0.027 ${M_{pl}}^2, x_{end}$ | 0.959 | 588 |
| 48.6 | $14M_{pl}$ | $(1/2 - 0.001/2\pi)f$ | $\pi - 0.001$ | $0.26{M_{pl}}^2$ | 0.959 | 63.1 |
| 49 | $14M_{pl}$ | $(1/2 - 0.001/2\pi)f$ | $\pi - 0.001$ | $1.30{M_{pl}}^2$ | 0.959 | 12.7 |
| 42 | $13M_{pl}$ | $(1/2 - 0.001/2\pi)f$ | $\pi - 0.001$ | $0.65{M_{pl}}^2$ | 0.953 | 25.3 |
| 36 | $12M_{pl}$ | $(1/2 - 0.001/2\pi)f$ | $\pi - 0.001$ | $1.30{M_{pl}}^2$ | 0.945 | 12.7 |

**Sample calculations (using number of e-folds)**

Equation (2.10) (number of e-folds) can be used to imposing data. If one uses this relation then the result will be

$$N = \frac{1}{M_{pl}^2}\int_{\varphi_{end}}^{\varphi_*}\frac{U}{U_{,\varphi}}d\varphi + \frac{1}{M_{pl}^2}\int_{x_{end}}^{x_*}\frac{V}{V_{,x}}dx \rightarrow$$

$$N \approx \frac{1}{4{M_{pl}}^2}[{\varphi_*}^2 - {\varphi_{end}}^2] + \frac{f^2}{4\pi^2{M_{pl}}^2}\ln\left(\frac{2}{1+\cos\frac{2\pi x}{f}}\right) \qquad (2.32)$$

**Note that equation (2.32) can be written more precisely in the following form.**

$$N = \frac{1}{4{M_{pl}}^2}[{\varphi_*}^2 - {\varphi_{end}}^2] + \frac{f^2}{4\pi^2{M_{pl}}^2}\ln\left(\frac{2}{1+\cos\frac{2\pi x_*}{f}}\right)$$

$$-\frac{f^2}{4\pi^2{M_{pl}}^2}\ln\left(\frac{2}{1+\cos\frac{2\pi x_{end}}{f}}\right) \qquad (2.32)'$$

**Note that one can ignore the last term in comparison to the other terms. The reason is, one needs axion field to be closest to the hilltop in the second term and in contrast demanding $\epsilon_x = 1$ for the end of inflation, therefore one can conclude $\alpha_{end} = \cos^{-1}\left(\frac{f^2 - 2\pi^2{M_{pl}}^2}{f^2 + 2\pi^2{M_{pl}}^2}\right)$, therefore setting $\cos\frac{2\pi x_{end}}{f} = 0$ is acceptable and the last term is ignorable**.

Now we need to obtain ${\varphi_{end}}^2$ in the first term. Considering single field inflation and remembering equation (2.12) with quadratic case we find

$$\epsilon_\varphi = \frac{{M_{pl}}^2}{2}\left(\frac{U_{,\varphi}}{U}\right)^2 = \frac{2M_{pl}^2}{\varphi^2} \qquad (2.12)$$

Inflation ends when $\epsilon_\varphi = 1$ so, ${\varphi_{end}}^2 = 2{M_{pl}}^2$.Therefore we have

$$N \approx \frac{1}{4{M_{pl}}^2}[{\varphi_*}^2 - 2{M_{pl}}^2] + \frac{f^2}{4\pi^2{M_{pl}}^2}\ln\left(\frac{2}{1+\cos\frac{2\pi x}{f}}\right) \qquad (2.33)$$

**The second part of the above equation is the same as Eq.(7) of authors in reference[11] with the exception of having one axion field here so no summation is needed**. Also first part of the equation is the same as single quadratic model of inflation. The rule is inserting various acceptable number of e-folds between 40-60 and verifying equation(2.33) for different values of axion decay constant $f$with initial conditions imposing on fields $\varphi_*, x_*$.Then evaluated values of $f$ are used in equations (2.25) and

(2.30).It is required to indicate that we use equation(2.25) at first in consistency with WMAP data for spectral index and then using such consistent initial conditions in equation (2.30) for $f_{nl}$.So, without imposing a constraint on $n-1$ we can achieve large $f_{nl}$.The aim is, obtaining large $f_{nl} \approx 20-25$ which is consistent with $n-1$.So,using $n-1 \approx 0$ in next steps it may be possible to see whether $f_{nl} \geq 20$ is still possible or not? Now we swap $\frac{1}{2}m^2\varphi^2$ to potential of the form $\lambda\varphi^4$ to see whether it is harder or easier to achieve large $f_{nl}$.So, the model is named $\lambda\varphi^4$ plus axion model.

### 3.$\lambda\varphi^4$ Plus axion model

In this case the model is given by following relation

$$W = \lambda\varphi^4 + \Lambda^4\left(1 - \cos\frac{2\pi x}{f}\right) \qquad (3.1)$$

Note that $\Lambda^4 = f^2m^2/4\pi^2$.So this relation can be written as

$$W = \lambda\varphi^4 + \frac{f^2m^2}{4\pi^2}\left(1 - \cos\frac{2\pi x}{f}\right) \qquad (3.2)$$

Using the same formulas one can find

$$U(\varphi) = \lambda\varphi^4\,, V(x) = \Lambda^4\left(1 - \cos\frac{2\pi x}{f}\right) \qquad (3.3)$$

$$\epsilon_\varphi = \frac{1}{M_{pl}^2}\left(\frac{U_{,\varphi}}{U}\right)^2 = \frac{8M_{pl}^2}{\varphi^2} \qquad (3.4)$$

$$\epsilon_x = \frac{M_{pl}^2}{2}\left(\frac{V_{,\varphi}}{V}\right)^2 = \frac{2\pi^2 {M_{pl}}^2}{f^2}\frac{1 + \cos\frac{2\pi x}{f}}{1 - \cos\frac{2\pi x}{f}} \qquad (3.5)$$

$$N_{,\varphi} = \frac{\partial N}{\partial\varphi} = \frac{1}{M_{pl}^2}\frac{U}{U_{,\varphi}} \to$$

$$N_{,\varphi} = \frac{\partial N}{\partial\varphi} = \frac{1}{M_{pl}^2}\frac{U}{U_{,\varphi}} = \frac{\varphi}{4M_{pl}^2} = \frac{1}{M_{pl}}\frac{1}{\sqrt{2\epsilon_\varphi}} \qquad (3.6)$$

$$N_{,\varphi\varphi} = \frac{\partial}{\partial\varphi}\left(\frac{\partial N}{\partial\varphi}\right) = \frac{1}{4M_{pl}^2} \qquad (3.7)$$

$$N_{,x} = \frac{\partial N}{\partial x} = \frac{1}{M_{pl}^2}\frac{V}{V_{,x}} = \frac{f}{2\pi M_{pl}^2}\frac{1 - \cos\frac{2\pi x}{f}}{\sin\frac{2\pi x}{f}} = \frac{1}{M_{pl}}\frac{1}{\sqrt{2\epsilon_x}} \qquad (3.8)$$

$$N_{,xx} = \frac{\partial}{\partial x}\left(\frac{\partial N}{\partial x}\right) = \frac{1}{M_{pl}^2}\frac{1}{1 + \cos\frac{2\pi x}{f}} \qquad (3.9)$$

$$N_{,\varphi}{}^2 = \frac{\varphi^2}{16M_{pl}{}^4} = \frac{1}{2M_{pl}^2}\frac{1}{\epsilon_\varphi} \qquad (3.10)$$

$$N_{,x}{}^2 = \frac{f^2}{4\pi^2 M_{pl}^4}\frac{1-\cos\frac{2\pi x}{f}}{1+\cos\frac{2\pi x}{f}} = \frac{1}{2M_{pl}^2}\frac{1}{\epsilon_x} \qquad (3.11)$$

Again we know that $N_{,x\varphi} = N_{,\varphi x} = 0$.So, observables like power spectrum, spectral index as well as $f_{nl}$(if detectable) can be computed now.It is required to indicate we use the same relations as before and just in the case of $\lambda\varphi^4$ the results are different from previous calculations but axion calculations remain unchanged.

**Power spectrum**

The power spectrum is given by (2.20)

$$p_\zeta = \frac{H_*{}^2}{4\pi^2}\sum_i N_{,i}N_{,i} = \frac{H_*{}^2}{4\pi^2}\left[N_{,x}{}^2 + N_{,\varphi}{}^2\right] \rightarrow$$

$$p_\zeta = \frac{H_*{}^2}{4\pi^2}\left[\frac{f^2}{4\pi^2 M_{pl}{}^4}\frac{1-\cos\left(\frac{2\pi x}{f}\right)}{1+\cos\left(\frac{2\pi x}{f}\right)} + \frac{\varphi^2}{16M_{pl}{}^4}\right] \qquad (3.12)$$

Using again $\epsilon_x, \epsilon_\varphi$ from equations (2.12) and (2.13) we can write power spectrum in terms of slow-roll parameters. It is found to be

$$p_\zeta = \frac{H_*{}^2}{8\pi^2 M_{pl}{}^2}\left[\frac{1}{\epsilon_x} + \frac{1}{\epsilon_\varphi}\right] \qquad (3.13)$$

Slow-roll conditions (2.8) implies that $p_\zeta$ can be written as

$$p_\zeta = \frac{W_*}{24\pi^2 M_{pl}{}^4}\left[\frac{1}{\epsilon_x} + \frac{1}{\epsilon_\varphi}\right] \qquad (3.14)$$

**SPECTRAL INDEX**

Spectral index is given by the same relation as follows:

$$n - 1 = -2\epsilon_H - \frac{1}{3H_*{}^2}\left[\frac{\frac{8\lambda\varphi^2}{\epsilon_\varphi} + \frac{8\pi^2\Lambda^4}{f^2}\frac{1}{\epsilon_x}}{\frac{1}{\epsilon_\varphi} + \frac{1}{\epsilon_x}}\right] \qquad (3.15)$$

Again if we substitute $\epsilon_x, \epsilon_\varphi$ then we can obtain a relation in terms of fields $x, \varphi$ which is useful for another numerical calculation. We find

$$n-1 = -2\epsilon_H - \frac{1}{3{H_*}^2}\left\{\left[\left(\frac{\frac{\lambda\varphi^4}{{M_{pl}}^2}}{\frac{\varphi^2}{8{M_{pl}}^2}+\frac{f^2}{2\pi^2{M_{pl}}^2}\left(\frac{1-\cos\frac{2\pi x}{f}}{1+\cos\frac{2\pi x}{f}}\right)}\right)\right] + \left[\frac{\frac{4\Lambda^4}{{M_{pl}}^2}\left(\frac{1-\cos\frac{2\pi x}{f}}{1+\cos\frac{2\pi x}{f}}\right)}{\frac{\varphi^2}{8{M_{pl}}^2}+\frac{f^2}{2\pi^2{M_{pl}}^2}\left(\frac{1-\cos\frac{2\pi x}{f}}{1+\cos\frac{2\pi x}{f}}\right)}\right]\right\} \quad (3.16)$$

We need to compute $\epsilon_H$and substitute into above equation. From its definition one can find

$$\epsilon_H = \frac{8\lambda^2\varphi^6{M_{pl}}^2 + \frac{f^2m^4{M_{pl}}^2\left(\sin\frac{2\pi x}{f}\right)^2}{8\pi^2}}{\left(\lambda\varphi^4+\Lambda^4\left(1-\cos\frac{2\pi x}{f}\right)\right)^2} \quad (3.17)$$

We are interested in numerical calculations, so we need to substitute (3.17) and ${H_*}^2$ from slow-roll equation into equation (3.16) in order to find an equation which is suitable for numerical calculations. We find

$$n-1 = -{M_{pl}}^2\left[\frac{16\lambda^2\varphi^6+\frac{f^2}{4\pi^2}m^4\left(\sin\frac{2\pi x}{f}\right)^2}{\left[\lambda\varphi^4+\frac{f^2m^2}{4\pi^2}\left(1-\cos\frac{2\pi x}{f}\right)\right]^2}\right] - \frac{{M_{pl}}^2}{\left[\lambda\varphi^4+\frac{f^2m^2}{4\pi^2}\left(1-\cos\frac{2\pi x}{f}\right)\right]}\left[\frac{\frac{f^2m^2}{\pi^2{M_{pl}}^2}\left(\frac{1-\cos\frac{2\pi x}{f}}{1+\cos\frac{2\pi x}{f}}\right)+\frac{\lambda\varphi^4}{{M_{pl}}^2}}{\frac{\varphi^2}{8{M_{pl}}^2}+\frac{f^2}{2\pi^2{M_{pl}}^2}\left(\frac{1-\cos\frac{2\pi x}{f}}{1+\cos\frac{2\pi x}{f}}\right)}\right] \quad (3.18)$$

**Tensor-scalar ratio**

Tensor to scalar ratio  is given by equation (1.26).It follows

$$r=\frac{p_T}{p_\zeta}=16\Big/\sum_i\frac{1}{\epsilon_i^*}=\frac{16}{\frac{1}{\epsilon_\varphi}+\frac{1}{\epsilon_x}} \tag{3.19}$$

**Calculation of parameter $\mathbf{f_{nl}}$**

If we compute $f_{nl}$from relation (2.28) we will get

$$f_{nl}\approx\frac{5}{6}\frac{\sum_{ij}N_{,i}N_{,j}N_{,ij}}{\left(\sum_k N_{,k}N_{,k}\right)^2}=\frac{5}{6}\frac{N_{,\varphi}{}^2N_{,\varphi\varphi}+N_{,x}{}^2N_{,xx}}{\left(N_{,\varphi}{}^2+N_{,x}{}^2\right)^2} \tag{2.28}$$

By substitution of $N_{,\ldots}$ we finally achieve following relation.

$$\frac{6}{5}f_{nl}\approx\frac{\frac{1}{2\epsilon_\varphi}+\frac{2}{\epsilon_x}\left(\frac{1}{1+\cos\frac{2\pi x}{f}}\right)}{\left(\frac{1}{\epsilon_\varphi}+\frac{1}{\epsilon_x}\right)^2} \tag{3.20}$$

In terms of fields $\varphi, x$ one can obtain following equation.

$$\epsilon_\varphi=\frac{8M_{pl}^2}{\varphi^2},\epsilon_x=\frac{1}{M_{pl}^2}\left(\frac{V_{,x}}{V}\right)^2=\frac{2M_{pl}^2\pi^2}{f^2}\left[\frac{1+\cos\frac{2\pi x}{f}}{1-\cos\frac{2\pi x}{f}}\right]\rightarrow$$

$$\frac{6}{5}f_{nl}\approx\frac{\frac{\varphi^2}{16M_{pl}{}^2}+\frac{f^2}{M_{pl}{}^2\pi^2}\left[\frac{1-\cos\frac{2\pi x}{f}}{1+\cos\frac{2\pi x}{f}}\right]\left[\frac{1}{1+\cos\frac{2\pi x}{f}}\right]}{\left[\frac{\varphi^2}{8M_{pl}{}^2}+\frac{f^2}{2M_{pl}{}^2\pi^2}\left(\frac{1-\cos\frac{2\pi x}{f}}{1+\cos\frac{2\pi x}{f}}\right)\right]^2} \tag{3.21}$$

In terms of tensor to scalar ratio $f_{nl}$ can be written as

$$\frac{6}{5}f_{nl}\approx\frac{r^2}{128}\left[\frac{1}{4\epsilon_\varphi}+\frac{1}{\epsilon_x}\left(\frac{1}{1+\cos\frac{2\pi x}{f}}\right)\right] \tag{3.22}$$

**NUMBER OF E-FOLDS**

Number of e-folds is given by equation (2.10).Therefore we find

$$N=\frac{1}{M_{pl}^2}\int_{\varphi_{end}}^{\varphi_*}\frac{U}{U_{,\varphi}}d\varphi+\frac{1}{M_{pl}^2}\int_{x_{end}}^{x_*}\frac{V}{V_{,x}}dx\rightarrow$$

$$N = \frac{1}{8M_{pl}{}^2}[\varphi_*{}^2 - \varphi_{end}{}^2] + \frac{f^2}{4\pi^2 M_{pl}{}^2}\ln\left(\frac{2}{1+\cos\frac{2\pi x}{f}}\right) \qquad (3.23)$$

Again $\varphi_{end}{}^2$ is found by requiring that inflation ends when $\epsilon_\varphi = 1$.So, $\varphi_{end}{}^2 = 8M_{pl}{}^2$. Then we find

$$N = \frac{1}{8M_{pl}{}^2}[\varphi_*{}^2 - 8M_{pl}{}^2] + \frac{f^2}{4\pi^2 M_{pl}{}^2}\ln\left(\frac{2}{1+\cos\frac{2\pi x}{f}}\right) \qquad (3.24)$$

**Numerical calculations**

For different values of $\varphi_*, x_*$ and $N$ provided that $N$ bounded between 40-60 we Investigate the results demanding consistency of initial conditions with $n \approx 0.96$ from WMAP. Because the form of equation (3.18) is complicated such that, we are not able to investigate spectral index when both fields dominate.In fact we are imposed to nominate either $\lambda\varphi^4$ domination or axion one ,so we are going to split $n$ into $n_\varphi, n_x$ which show spectral index for $\lambda\varphi^4$ and axion domination respectively. But from the specific form of relation (3.18) **one can say that axion domination is impossible because in this model we have just one axion around hilltop so acceptable spectral index in this case is not achievable**. Frankly speaking we can't use such approximation. **As a result; we are ignoring this case in the following table. The results are outlined in table 2.**

Table (2)

| $N$ | $\varphi_*$ | $x_*$ | $\alpha_*$ | $f^2$ | $n_\varphi$ | $f_{nl}$ |
|---|---|---|---|---|---|---|
| 50 | $12M_{pl}$ | $(1/2 - 0.001/2\pi)f$ | $\pi - 0.001$ | $85.71M_{pl}{}^2$ | 0.888 | 0.19 |
| 50 | $14M_{pl}$ | $(1/2 - 0.001/2\pi)f$ | $\pi - 0.001$ | $68.83M_{pl}{}^2$ | 0.918 | 0.24 |
| 50 | $16M_{pl}$ | $(1/2 - 0.001/2\pi)f$ | $\pi - 0.001$ | $49.35M_{pl}{}^2$ | 0.938 | 0.33 |
| 50 | $18M_{pl}$ | $(1/2 - 0.001/2\pi)f$ | $\pi - 0.001$ | $27.27M_{pl}{}^2$ | 0.951 | 0.60 |
| 50 | $20M_{pl}$ | $(1/2 - 0.001/2\pi)f$ | $\pi - 0.001$ | $2.60M_{pl}{}^2$ | 0.959 | 6.31 |
| 60 | $12M_{pl}$ | $(1/2 - 0.001/2\pi)f$ | $\pi - 0.001$ | $111.69M_{pl}{}^2$ | 0.888 | 0.15 |
| 60 | $16M_{pl}$ | $(1/2 - 0.001/2\pi)f$ | $\pi - 0.001$ | $75.32M_{pl}{}^2$ | 0.938 | 0.22 |
| 60 | $22M_{pl}$ | $(1/2 - 0.001/2\pi)f$ | $\pi - 0.001$ | $1.30M_{pl}{}^2$ | 0.967 | 12.7 |
| 59.6 | $22M_{pl}$ | $(1/2 - 0.001/2\pi)f$ | $\pi - 0.001$ | $0.26M_{pl}{}^2$ | 0.967 | 63.2 |
| 59.6 | $22M_{pl}$ | $(1/2-0.001/2\pi)f$ | $\pi-0.001$ | $0.27{M_{pl}}^2$, $x_{end}$ | 0.967 | 60.8 |
| 49.5 | $20M_{pl}$ | $(1/2-0.001/2\pi)f$ | $\pi-0.001$ | $1.30{M_{pl}}^2$ | 0.959 | 12.7 |
| 49.1 | $20M_{pl}$ | $(1/2-0.001/2\pi)f$ | $\pi-0.001$ | $0.26{M_{pl}}^2$ | 0.959 | 63.2 |

**Verification of large $f_{nl}$ by imposing $n-1\approx 0$**

We are now verifying if one sets spectral index almost one (i.e.$n-1\approx 0$, it may be possible to see whether $f_{nl}\gtrsim 20$ or not.

The idea is to set $n-1=0$ in equations (2.25) and (3.18).Let us start from equation (2.25).We have

$$\frac{\left[m^4\varphi^2{M_{pl}}^2+\frac{f^2m^4{M_{pl}}^2}{4\pi^2}\left(\sin\frac{2\pi x}{f}\right)^2\right]}{\left[\frac{1}{2}m^2\varphi^2+\frac{f^2m^2}{4\pi^2}\left(1-\cos\frac{2\pi x}{f}\right)\right]^2}$$
$$=-\frac{\left[\frac{f^2m^2}{\pi^2}\left(\frac{\left(\sin\frac{2\pi x}{f}\right)^2}{\left(1+\cos\frac{2\pi x}{f}\right)^2}\right)+m^2\varphi^2\right]}{\left[\frac{1}{2}m^2\varphi^2+\frac{f^2m^2}{4\pi^2}\left(1-\cos\frac{2\pi x}{f}\right)\right]}\left(\frac{1}{\frac{f^2}{2\pi^2{M_{pl}}^2}\left(\frac{1-\cos\frac{2\pi x}{f}}{1+\cos\frac{2\pi x}{f}}\right)+\frac{\varphi^2}{2{M_{pl}}^2}}\right)$$

(3.25)

After simplification we get

$$\left\{\frac{f^2}{\pi^2}\left(\frac{1-\cos\frac{2\pi x}{f}}{1+\cos\frac{2\pi x}{f}}\right)+\varphi^2\right\}\left\{\frac{1}{2}\left[\varphi^2+\frac{f^2}{4\pi^2}\sin^2\left(\frac{2\pi x}{f}\right)\right]\right.$$
$$\left.+\left[\frac{1}{2}\varphi^2+\frac{f^2}{4\pi^2}\left(1-\cos\frac{2\pi x}{f}\right)\right]\right\}=0 \qquad (3.26)$$

Therefore one finds two solutions;

$$\varphi^2=-\frac{f^2}{\pi^2}\left(\frac{1-\cos\frac{2\pi x}{f}}{1+\cos\frac{2\pi x}{f}}\right) \qquad (3.27)$$

And

$$\varphi^2=-\frac{f^2}{4\pi^2}\left\{\frac{1}{2}\sin^2\left(\frac{2\pi x}{f}\right)+\left(1-\cos\frac{2\pi x}{f}\right)\right\} \qquad (3.28)$$

**Both of solutions are imaginary**. There are two options for us. First is to set $\varphi = 0$ as a solution into relevant relation(2.30) and second substituting these solutions into equation(2.30) and verifying the possibility of field being complex. Therefore, One can find that the first solution is impossible (equation (3.27)) because of singularity. For the second solution one finds following relation.

$$\frac{6}{5}f_{nl} \approx \frac{4{M_{pl}}^2\pi^2}{f^2}\frac{-\frac{1}{16}\sin^2\left(\frac{2\pi x}{f}\right) - \frac{1}{8}\left(1-\cos\frac{2\pi x}{f}\right) + \frac{1-\cos\left(\frac{2\pi x}{f}\right)}{\left(1+\cos\left(\frac{2\pi x}{f}\right)\right)^2}}{\left[-\frac{1}{8}\sin^2\left(\frac{2\pi x}{f}\right) - \frac{1}{4}\left(1-\cos\frac{2\pi x}{f}\right) + \frac{1-\cos\left(\frac{2\pi x}{f}\right)}{1+\cos\left(\frac{2\pi x}{f}\right)}\right]^2} \quad (3.29)$$

We can ignore some terms like first and second terms both in numerator and denominator provided that we set $\alpha = \frac{2\pi x}{f}$ very near to the hilltop. Therefore simplification gives us following relation. **Note that by inserting the value of $\boldsymbol{\varphi}$ which is imaginary one can say that the field is not real valued field, but in next step one sees ignoring the value of $\boldsymbol{\varphi}$ both in numerator and denominator is translated to be $\boldsymbol{\varphi}$ as a real valued field and the only possibility for $\boldsymbol{\varphi}$ to be real is to set $\boldsymbol{\varphi = 0}$ as a solution to the condition $\boldsymbol{n - 1 \approx 0}$.Therefore we can argue that, both fields are real fields and there is nothing wrong with our verification. In addition, setting $\boldsymbol{\varphi = 0}$ in our model reduces it to a single inflation field model. Then we find**

$$\frac{6}{5}f_{nl} \approx \frac{4{M_{pl}}^2\pi^2}{f^2}\frac{1}{1-\cos\left(\frac{2\pi x}{f}\right)} \quad (3.30)$$

Using fact that very near the hilltop one can estimate $\alpha = \frac{2\pi x}{f} = \pi - 0.001$ then equation (3.30) is converted to the following relation.

$$\frac{6}{5}f_{nl} \approx \frac{2{M_{pl}}^2\pi^2}{f^2} \quad (3.31)$$

Then

$$f_{nl} \approx \frac{5\pi^2}{3}\left(\frac{M_{pl}}{f}\right)^2 \quad (3.32)$$

**The value of $\boldsymbol{f_{nl}}$ in above equation is large enough to be detected. For instance setting $\boldsymbol{f = M_{pl}}$ gives us $\boldsymbol{f_{nl} \approx 16.4}$.For $\boldsymbol{f = 0.4M_{pl}}$ we have $\boldsymbol{f_{nl} \approx 103}$.It is** **important to consider the fact that the value of $f_{nl}$ in this case is based on single inflation model ($\varphi = 0$).**

For the time being imposing $n - 1 \approx 0$ in equation (3.18) ($\lambda\varphi^4$) case does not give us any significant relation.

**Conclusions**

What we obtain for $f_{nl}$ in some cases satisfy both requirements. First consistency with spectral index from WMAP and second large non-gaussianity of order 10 and even 100 or of order 1000.Of course, there is a constraint on obtaining large $f_{nl}$.**The problem is, if we decrease $f$ in order to achieve large value for $f_{nl}$ then we need to be careful about number of e-folds. It is said that number of e-folds must be between $40 - 60$**.However, relation (3.24) shows how it behaves. When we fix $\alpha_* = \pi - 0.001 \rightarrow x_* = (1/2 - 0.001/2\pi)f$ it means the only possibility to increase $f_{nl}$ is to change initial condition on value of $\varphi_*$ conditional on number of e-folds remains in acceptable range(i.e. 40-60).As a result there is an upper bound for $\varphi_* = 22M_{pl}$ with $N = 60$ generates $f_{nl} \approx 12.7$.But ,similarly if one can decrease $f$ then can produce large $f_{nl} \approx 63.2$ with $f^2 = 0.26{M_{pl}}^2$ in $\lambda\varphi^4$.**Accordingly, we can achieve $f_{nl} \approx 607$** in first case(quadratic+ axion) with the requirements of $f^2 = 0.026{M_{pl}}^2$and $\varphi_* = 22M_{pl}$ with $N = 48.51$.It seems we can achieve any large $f_{nl}$ that we are interested in. But that is not true. In fact, strong constraint comes from our initial condition on axion field $x_* = (1/2 - 0.001/2\pi)f = 0.5f$.So, if one chooses $f^2 = 0.026{M_{pl}}^2$ then it reads $x_* = 0.5f = 0.08M_{pl}$ .Therefore one can't make any large $f_{nl}$.In axion N-flation models(reference[16]) the constraint is, if one wants to decrease $f$ then it requires to increase number of fields because in that paper Soo A. Kim, Andrew Liddle and David Seery have shown that the number of e-folds scales with $\left(\frac{f_i}{M_{pl}}\right)^2$number of fields(relation 7), so if one decreases $f$ then one has to increase the number of fields to compensate *quadratically*.In our model, the problem is, most of axions are replaced by a quadratic and then by $\lambda\varphi^4$ potentials. So, we need to have an axion around the hilltop. Therefore reasonably, we fix $\alpha_* = \pi - 0.001$ , therefore $x_*$ being so that one axion stay around the hilltop. Then the only manoeuvre is to change $\varphi_*$ and $N$ with the requirement of $x_*$ not being less than $0.001M_{pl}$[15].Recall that as one decreases $f$ then makes the fluctuations stronger and one can potentially enter a regime where they dominate the classical rolling. In this regime all the usual methods of making predictions break down.

**Imposing $n - 1 \approx 0$**

**Imposing the above approximation reduces our model to a single field inflation model which gives us large $f_{nl}$ of order 10 and 100 and even 1000, depending on the value of $f$(see equation (3.32).This conclusion is in contradiction of saying that any detection of large $f_{nl}$ will rule out single inflation models (see reference [5]).**